\documentclass[12pt]{iopart}
\usepackage{iopams}

\begin{document}

\title[Thermodynamic properties of the $2+1$-dimensional Dirac fermions]{Thermodynamic properties of the $2+1$-dimensional
Dirac fermions with broken time-reversal symmetry}

\author{S G Sharapov}

\address{Bogolyubov Institute for Theoretical Physics, National Academy of Science of Ukraine, 14-b
        Metrologicheskaya Street, Kiev 03680, Ukraine}

\ead{sharapov@bitp.kiev.ua}
\vspace{10pt}

%\begin{indented}
%\item[]March 2015
%\end{indented}

\begin{abstract}
We study the thermodynamic properties of the two-component $2+1$-dimensional massive Dirac fermions
in an external magnetic field. The broken time-reversal symmetry results
in the presence of a linear in the magnetic field  part of the thermodynamic potential, while in the famous problem of
Landau diamagnetism the leading field dependent term is quadratic in the field. Accordingly,
the leading term of the explicitly calculated magnetization is anomalous, viz. it is independent of the strength
of the magnetic field.  The  St\v{r}eda formula is employed to describe how the anomalous magnetization is related to
the anomalous Hall effect.
\end{abstract}

%\keywords{Landau levels, Dirac fermions, magnetic moment}
\submitto{\jpa}

\section{Introduction and Model}
\label{sec:intro}

Since  80s of the 20th century the condensed matter realization of the $(2+1)$ - dimensional Dirac fermions
\cite{Semenoff1984PRL,Haldane1988PRL} attracts attention of researchers. The discovery of graphene in 2004 \cite{Geim2007NatMat}
was a tremendous breakthrough in the experimental realization of the massless Dirac fermions. Now a family of the solid-state
Dirac materials which includes both the  brothers of graphene, viz. made in 2010 silicene, announced in 2014
 germanene \cite{Davila2014NJP}, and cousins such as MoS$_2$,
topological insulators, Weyl semimetals, etc. There are even artificially designed in 2012 nephews such as molecular graphene \cite{Gomes2012Nature} and ultracold fermionic atoms in optical honeycomb lattice \cite{Tarruell2012Nature}.
An exciting feature of the latter artificial condensed matter system is that it allows one the experimental realization
\cite{Jotzu2014Nature} of the Haldane model \cite{Haldane1988PRL} where the time-reversal symmetry is broken and
a quantum Hall effect appears as an intrinsic property of the band structure in the {\it absence\/} of an external magnetic field.

As a matter of fact, the low-energy description of these systems is based
on a pair of the independent effective Dirac Hamiltonians
\begin{equation}
\label{Hamiltonian}
\mathcal{H}_\eta(\mathbf{k}) = \hbar v_F (\eta k_x \tau_1 + k_y \tau_2) + \Delta_\eta \tau_3,
\end{equation}
where the Pauli matrices $\pmb{\tau}$ act in the sublattice (pseudospin) space, the two-dimensional wave-vector
$\mathbf{k} = (k_x, k_y) $ is counted off the two independent $\mathbf{K}_{\eta}$ points (with $\eta = \pm$) in
the Brillouin zone,  and $v_F$ is the Fermi velocity, and $\Delta_\eta$ is the mass (gap) term.
Although the spin degree of freedom is neglected for simplicity, it can be easily included if necessary. Accordingly,
the most general expression for the gap
is given by
\begin{equation}
\Delta_\eta = \Delta + \eta \Delta_T,
\end{equation}
where the gap $\Delta$ is invariant with respect to the time-reversal symmetry, and $\eta \Delta_T$ is not.

In the simplest case \cite{Semenoff1984PRL},
$\Delta_T =0$, the mass term $\Delta$ originates from a different on-site energy on
two triangular sublattices that form hexagonal lattice. The fermion doubling \cite{Nielsen1981NPB}
guarantees that the ordinary solid-state Dirac materials are always described by pairs of the Hamiltonians (\ref{Hamiltonian}),
so that the full Hamiltonian $\mathcal{H}(\mathbf{k}) = \mathcal{H}_{\eta=+1}(\mathbf{k}) \oplus \mathcal{H}_{\eta=-1}(\mathbf{k}) $
respects time-reversal symmetry
(see Ref.~\cite{Gusynin2007IJMPB} for an overview of the discrete symmetries for the various mass terms),
\begin{equation}
(\Pi \otimes  \tau_0 ) \mathcal{H}^\ast(\mathbf{k}) (\Pi \otimes \tau_0 ) =
\mathcal{H}(-\mathbf{k})
\end{equation}
that involves operator $\Pi$ swapping $\eta=1$ and $\eta =-1$ valleys. Here $\tau_0$ is the unit matrix.
The inversion symmetry is, however, broken
because the sublattices are inequivalent for $\Delta \neq 0$. Obviously, when a separate $\mathbf{K}_\eta$ point is considered,
the time-reversal symmetry is always broken.

A more sophisticated case, $\Delta_{T} \neq 0$,
with broken time-reversal symmetry is realized in the Haldane model \cite{Haldane1988PRL} by including
next-nearest-neighbour (nearest-neighbour on the same sublattice) hopping and periodic local magnetic flux
density with zero total flux through the unit cell. The experimental implementation of this model in \cite{Jotzu2014Nature}
is highly nontrivial. An earlier history of searches of the condensed matter realization of the time-reversal anomaly is
presented in \cite{Tchernyshyov2000PRB}.

From a theoretical point of view, the specifics of the Haldane model appears to be more transparent when
one considers the spectrum of the Hamiltonians (\ref{Hamiltonian}) in an external magnetic field
$\mathbf{B} = \mathbf{\nabla} \times \mathbf{A} = (0,0,B)$ applied perpendicular
to the plane along the positive $z$ axis. Accordingly, the momentum operator $\hat{p}_i = - i \hbar \partial_i$
has to be replaced by the covariant momentum $\hat{p}_i \to \hat{p}_i + \frac{e}{c} A_i$. Here $-e<0$ is the electron charge
and $\mathbf{A} = (0, Bx,0)$ is the vector potential in the Landau gauge. The corresponding
Landau level energies are
\begin{equation}
\label{spectrum}
\epsilon_{n\eta}=
\cases{
-\eta \Delta_\eta \mbox{sgn} (eB), & n=0, \\
\pm M_n, & n= 1,  2, \dots, \\}
\end{equation}
where
$M_n = \sqrt{\Delta_\eta^2+2 n v^2_{\scriptscriptstyle F}\hbar |eB|/ c}$.

In $2 + 1$ dimensions there are two inequivalent irreducible $2 \times 2$
representations of the Dirac algebra labelled by $a=1,2$. In particular, the Hamiltonian (\ref{Hamiltonian})
corresponds to the representation $\gamma_a^\nu = (\tau_3, - i \eta \tau_2, i \tau_1)$ with $\nu=0,1,2$.
In general, one can relate the sign of the energy of the $n=0$ Landau level to the sign of the product of the mass,
$\Delta_\eta$, and the signature, $\eta_a$, of the Dirac matrices $\gamma^\nu_a$.
The signature is defined as \cite{Hosotani1993PLB}
\begin{equation}
\eta_a = \frac{i}{2} \mbox{tr} [\gamma^0_a \gamma^1_a \gamma^2_a] = \pm 1
\end{equation}
and is called sometimes ``chirality''.
Thus, instead of putting the $\pm$ sign in the matrix $\gamma^1_a$,
one can instead include this sign in the mass term (see e.g. Refs.~\cite{Semenoff1984PRL,Andresen1995PRD}).

As mentioned before, in the time-reversal symmetric case, $\Delta_T=0$, the spectrum of the full
Hamiltonian $\mathcal{H}(\mathbf{k}) = \mathcal{H}_{\eta=+1}(\mathbf{k}) \oplus \mathcal{H}_{\eta=-1}(\mathbf{k})$
is particle-hole symmetric and invariant under $B \to -B$.
The case $\Delta_T \neq0$ and $\Delta = 0$   corresponds to the
so called Haldane  mass that breaks time-reversal symmetry. As the result, the $n=0$ ``zero-mode'' level breaks both the
particle-hole symmetry and invariance under $B \to -B$. It is crucial that the difference between these cases
survives even in the  $B=0$ limit, making possible the quantum Hall effect in the absence of magnetic field.

The vast majority  of the literature (see e.g. Refs.~\cite{Cangemi1996AP,Sharapov2004PRB,Tabert2015PRB} and
references therein) on the thermodynamic properties of the Dirac fermions including magnetic oscillations
is devoted to the case of the particle-hole symmetric spectrum. This concerns both the papers
that consider the reducible $4 \times 4$ representation of the Dirac algebra in $2 + 1$ dimensions
and present a field theoretical perspective of the problem \cite{Cangemi1996AP}
and the works  devoted to the condensed matter systems \cite{Sharapov2004PRB,Tabert2015PRB}
with even number of fermion species.
Since the time-reversal symmetry is preserved,
a finite Hall conductivity is only possible in nonzero external magnetic field.
Respectively, the first magnetic field dependent term of the grand thermodynamic
potential $\Omega(\mu,B)$ is proportional  to $B^2$, so that there is no net magnetization in the limit $B \to 0$.

The purpose of the present work is to study the thermodynamic properties of the Dirac fermions with the
broken time-reversal symmetry.
In other words, we focus on the asymmetric spectrum (\ref{spectrum}) associated with the Hamiltonian
(\ref{Hamiltonian}) for one fixed value of $\eta$, viz. we take $\Delta_T =0$, and have
$\Delta_\eta = \Delta$.
Having the thermodynamic potential for ``unphysical'' case with a fixed $\eta$, it is straightforward to
obtain, for example, the result for two $\mathbf{K}_{\pm}$ points with the Haldane mass and
include the spin degree of freedom.

The paper is organized as follows. We already began by presenting
in Sec.~\ref{sec:intro} the problem and the model Hamiltonian (\ref{Hamiltonian}).
In Sec.~\ref{sec:Grand-Omega} we consider the grand thermodynamic potential.
We show that the unboundedness of the Landau level spectrum  for the negative energies results
in the diverging carrier density and discuss the relativistic form of the thermodynamic potential
that corresponds to the finite carrier imbalance. The magnetic field dependent parts
of the carrier imbalance and density  are obtained in
Sec.~\ref{sec:density} using the Euler-Maclaurin formula. The
main results of the present paper that include  thermodynamic potential
and magnetization in the low magnetic field regime are presented in
Sec.~\ref{sec:magnetization}.  Using the St\v{r}eda formula we overview in Sec.~\ref{sec:Hall}
how the obtained expressions can be linked to the known results for the Hall conductivity in the anomalous and normal cases.
In Sec.~\ref{sec:concl}, the main results of the paper are summarized.

\section{Grand thermodynamic potential and its specific in the Dirac case}
\label{sec:Grand-Omega}

The grand thermodynamic potential \cite{Landau5:book} can be written as follows
\begin{equation}
\label{Omega.nonrel-def}
\Omega(T,\mu) = - T \int\limits_{-\infty}^{\infty} d \epsilon
D(\epsilon) \ln \left(1 + e^{\frac{\mu - \epsilon}{T}} \right),
\end{equation}
where $T$ is the temperature, $\mu$ is the chemical potential, and $D(\epsilon)$ is the density of states (DOS).
We have also  set the Boltzmann constant $k_B=1$.
The derivative of the thermodynamic potential (\ref{Omega.nonrel-def}) with
respect to the chemical potential $\mu$,
\begin{equation}
\label{number-nonrel0}
\rho (T,\mu)= - \frac{1}{V} \frac{\partial \Omega(T,\mu)}{\partial \mu} =
\int\limits_{-\infty}^\infty d\epsilon
D(\epsilon) n_F(\epsilon),
\end{equation}
determines the density of carriers $\rho$ in nonrelativistic many-body
theory as a function of $T$, $B$, and $\mu$.
Here $n_F(\epsilon) = 1/[\exp[(\epsilon-\mu)/T] +1]$ is the usual
Fermi function and $V$ is volume (area) of the system which is set to be unit.

%On the other hand, one can consider Eq.  as an equation for m as a function
%of T, B, and n, which is typical for studies in a canonical ensemble.

In the absence of scattering from impurities
the DOS per unit area for a given value of $\eta$ is expressed via
the energies of the Landau levels $\epsilon_{n\eta}$ as follows
\begin{equation}
\label{DOS}
D_0(\epsilon)  =  \frac{|eB|}{2 \pi \hbar c} \Big[\delta(\epsilon + \eta \Delta \mbox{sgn} (eB))
 + \sum_{n=1}^{\infty} [\delta(\epsilon - M_n) + \delta(\epsilon + M_n)] \Big].
\end{equation}
Substituting the DOS $D_0(\epsilon)$ given by Eq.~(\ref{DOS}) in Eq.~(\ref{number-nonrel0}) we obtain
\begin{equation}
\label{number-nonrel-1}
\fl
\rho \equiv  \rho_0 + \rho_{n \geq 1}= \frac{|eB|}{ 2 \pi \hbar c} \Bigg[ n_F(- \eta \Delta \mathrm{sign} (eB) )
\left. + \sum_{n=1}^{\infty} [n_F(M_n ) + n_F(-M_n )]
\right],
\end{equation}
where $\rho_0 $ and $\rho_{n \geq 1}$ denote contributions from the lowest, $n=0$, and remaining ($\pm M_n$, $n \neq 0$)
Landau levels, respectively.
We observe that Eq.~(\ref{number-nonrel-1}) diverges because the spectrum (\ref{spectrum}) is unbound for the
negative energies $\epsilon_{n\eta}$.  Since the Dirac Hamiltonian (\ref{Hamiltonian})
has to be regarded as an effective model Hamiltonian derived from the tight-binding model,
the divergence can be removed by using the appropriate band-width cutoff.

The other way to remove this divergency is to begin with so called relativistic thermodynamic potential
(see Refs.~\cite{Cangemi1996AP,Niemi1985NPB,Gusynin1995PRD})
\begin{equation}
\label{Omega.rel-def}
\Omega^{\mathrm{rel}}(T,\mu)=-T\int\limits_{-\infty}^\infty d\epsilon
D(\epsilon)\ln\left( 2\cosh\frac{\epsilon-\mu}{2T}\right) .
\end{equation}
To clarify the physical meaning of the potential (\ref{Omega.rel-def}) we differentiate it with
respect to $\mu$ and obtain the relativistic carrier density
\begin{equation}
\label{number-rel}
\rho^{\mathrm{rel}}(T,\mu) = - \frac{\partial \Omega^{\mathrm{rel}}(T,\mu)}{\partial \mu} =
-\frac{1}{2} \int\limits_{-\infty}^\infty d\epsilon D(\epsilon) \tanh \frac{\epsilon - \mu}{2T}.
\end{equation}
Then for the particle-hole symmetric spectrum, i.e. when the DOS $D(\epsilon)$ is an even function of the energy $\epsilon$,
$D(\epsilon) = D (- \epsilon)$ one can see that $\rho^{\mathrm{rel}}(T,\mu=0) =0$. Further,
using the identity $1 = \theta(\epsilon) + \theta(-\epsilon)$ with $\theta(\epsilon)$ being a step function,
one can rewrite the last equation in the following form
\begin{equation}
\rho^{\mathrm{rel}} = \int\limits_{-\infty}^\infty d\epsilon D(\epsilon) \left[ n_F(\epsilon) \theta(\epsilon) -
[1 - n_F(\epsilon)] \theta(- \epsilon)
\right].
\end{equation}
This shows that $\rho^{\mathrm{rel}}$ corresponds to the relativistic carrier density or
the carrier imbalance, viz. $\rho^{\mathrm{rel}} = \rho_+ - \rho_-$,
where $\rho_+$ and $\rho_-$ are the densities of ‘‘electrons’’ and ‘‘holes’’, respectively.
Accordingly, when the DOS is the even function of energy, the relativistic thermodynamic potential (\ref{Omega.rel-def})
can be presented as a sum of the vacuum, electron and hole terms \cite{Cangemi1996AP}
illustrating its physical meaning. The interpretation of $\rho^{\mathrm{rel}} $ for the case of the
asymmetric DOS is discussed in detail in \cite{Niemi1985NPB} (see also Ref.~\cite{Niemi1986PR}  for a review).

Substituting the DOS, $D_0(\epsilon)$ given by Eq.~(\ref{DOS}) in Eq.~(\ref{number-rel}),  one obtains
\begin{equation}
\label{number-rel1}
\fl
\rho^{\mathrm{rel}} \equiv  \rho^{\mathrm{rel}}_0 + \rho^{\mathrm{rel}}_{n \geq 1}=
\frac{|eB|}{ 4 \pi \hbar c} \Bigg[ \tanh \frac{\mu + \eta \Delta \mathrm{sign} (eB)}{2T}
 +
\sum_{n=1}^{\infty} \left[\tanh \frac{\mu - M_n}{2T} + \tanh \frac{\mu + M_n}{2T} \right] \Bigg],
\end{equation}
where $\rho^{\mathrm{rel}}_0 $ and $\rho^{\mathrm{rel}}_{n \geq 1}$ denote contributions from the lowest and remaining
Landau levels, respectively. One can see that, in contrast to Eq.~(\ref{number-nonrel-1}),
Eq.~(\ref{number-rel1}) converges. Thus the use of the relativistic potential (\ref{Omega.rel-def}) makes the calculation
free of divergencies, but at the end one should prove that the physical results do not change
as compared to the results that follow directly from the initial potential  (\ref{Omega.nonrel-def}).

In general, one can relate the grand thermodynamic potential (\ref{Omega.nonrel-def}) and the
relativistic potential  (\ref{Omega.rel-def}) as follows
\begin{equation}
\label{Omega-linked}
\Omega (T,\mu) = \Omega^{\mathrm{rel}}(T,\mu)
- \frac{1}{2} \int_{-\infty}^{\infty} d\epsilon (\mu - \epsilon) D(\epsilon).
\end{equation}
When the $D(\epsilon)$ is an even function of the energy $\epsilon$, the last term
with the integral in Eq.~(\ref{Omega-linked}) reduces to
\begin{equation}
\label{Omega-linked-symmetric}
- \mu \rho_{\mathrm{hf}} = - \mu \int_{-\infty}^{\infty} d\epsilon D(\epsilon) \theta(- \epsilon).
\end{equation}
Then the carrier imbalance $\rho^{\mathrm{rel}}$ and total carrier density $\rho$
are related by the expression
$\rho^{\mathrm{rel}} (T,\mu) = \rho (T,\mu) - \rho_{\mathrm{hf}}$, where $\rho_{\mathrm{hf}}$ is the density of particles
for a half-filled band (in the lower Dirac cone). Note that both $\rho (T,\mu)$ and  $\rho_{\mathrm{hf}}$
are divergent, so that the appropriate cutoff has to be introduced. This reflects the already mentioned fact that
in the continuum field theory there is no lower bound to the Dirac sea of filled electron states.

The thermodynamic potential (\ref{Omega.rel-def}) can be restored from the
particle density  $\rho^{\mathrm{rel}} (\mu,T,B)$ by integration
\begin{equation}
\label{Omega-integrate}
\Omega^{\mathrm{rel}} (T,\mu,B) = - \int_{-\infty}^{\mu} d \epsilon \rho^{\mathrm{rel}} (T,\epsilon,B) + \Omega_c(T,B).
\end{equation}
Here $\Omega_c(T,B)$ is a constant of integration.
Similarly, integrating $\rho(\mu,T,B)$ one restores the potential  (\ref{Omega.nonrel-def}).

The magnetization $M$ in the direction perpendicular to the
plane  is defined in the grand canonical ensemble by the derivative with respect to $B$ at fixed chemical
potential, i.e.
\begin{equation}
\label{magnetization-def}
M (T,\mu,B) = - \frac{\partial \Omega (T,\mu,B)}{\partial B}.
\end{equation}
Thus one needs only the magnetic field dependent part of the grand thermodynamic
potentials (\ref{Omega.rel-def}) and (\ref{Omega.nonrel-def}). A special interest for us represents
a linear in magnetic field part of the thermodynamic potential those presence would imply
a nonzero net magnetization in the limit of $B \to 0$.

\section{Calculation of the carrier imbalance and density}
\label{sec:density}

Since we are interested in the weak field regime, there is no need to evaluate the sum over Landau levels
exactly. The simplest way to extract the magnetic field dependent terms  is to use the Euler-Maclaurin formula
\begin{equation}
\label{E-M}
\frac{1}{2} F(0) + \sum_{n=1}^{\infty} F(n) \approx \int_0^\infty F(x) d x - \frac{1}{12} F^\prime (0)
\end{equation}
following the seminal paper on Landau diamagnetism \cite{Landau1930ZPhys}. Firstly
we calculate the carrier imbalance $\rho^{\mathrm{rel}}_{n \geq 1}$ from Eq.~(\ref{number-rel1}), so that
\begin{equation}
\label{terms-n>1}
\rho^{\mathrm{rel}}_{n \geq 1} = \frac{|eB|}{ 4 \pi \hbar c} \sum_{n=1}^\infty F^{\mathrm{rel}}(n)
\end{equation}
with
\begin{equation}
F^{\mathrm{rel}}(n) = \tanh \frac{\mu - M_n}{2T} + \tanh \frac{\mu + M_n}{2T}.
\end{equation}
Repeating the arguments of Refs.~\cite{Landau1930ZPhys,Landau5:book} one can see that the use of the
Euler-Maclaurin formula is justified in
the weak-field regime, $L(B) \ll |\mu|$ with $L(B) = \sqrt{2 |eB| \hbar v_F^2/c}$ being the Landau scale.

One can see that the term with the integral in Eq.~(\ref{E-M}) is independent
of the magnetic field similarly to the original Landau consideration \cite{Landau1930ZPhys}.
Combining together the lowest Landau contribution,
$\rho^{\mathrm{rel}}_0$ and term containing $(-1/2) F^{\mathrm{rel}}(0)$, we obtain
the linear in $B$ contribution, $\rho^{\mathrm{rel}}_{\mathrm{I}}$, to $\rho^{\mathrm{rel}}$:
\begin{equation}
\label{number-linear-B}
\fl
\rho^{\mathrm{rel}}_{\mathrm{I}}  =  \frac{|e B|}{4 \pi \hbar c}
\left[ \tanh \frac{\mu + \eta \Delta \mathrm{sign} (eB)}{2T} \right.
\left. - \frac{1}{2} \tanh \frac{\mu - |\Delta|}{2T} - \frac{1}{2} \tanh \frac{\mu + |\Delta|}{2T}
\right].
\end{equation}
Analyzing all possible cases $e B, \eta \Delta  \gtrless 0$,
one can simplify the last expression to the form
\begin{equation}
\label{number-linear-B-fin}
\rho^{\mathrm{rel}}_{\mathrm{I}} = \frac{eB \mbox{sgn} (\eta \Delta)}{8 \pi \hbar c} \left[
\tanh \frac{\mu + |\Delta|}{2T} - \tanh \frac{\mu - |\Delta|}{2T}
\right].
\end{equation}
Note that there is no linear in $B$ term in the analysis of Landau diamagnetism
\cite{Landau1930ZPhys,Landau5:book}.  This term would also be absent if one considers the total
contribution from both valleys with $\eta = \pm$.
Similarly to \cite{Landau1930ZPhys,Landau5:book}, the term with the derivative
$(F^{\mathrm{rel}})^\prime(0)$ produces the quadratic in $B$ part, $\rho^{\mathrm{rel}}_{\mathrm{II}}$, of $\rho^{\mathrm{rel}}$:
\begin{equation}
\label{number-quadratic-B-fin}
\rho^{\mathrm{rel}}_{\mathrm{II}}   = -\frac{(eB)^2  v_F^2}{96 \pi c^2 |\Delta| T}
\left[\frac{1}{\cosh^2 \frac{\mu + |\Delta|}{2T}} - \frac{1}{\cosh^2 \frac{\mu - |\Delta|}{2T}}   \right].
\end{equation}

The same approach can be applied to the carrier density (\ref{number-nonrel-1}). However,
as mentioned earlier, the ``nonrelativistic'' analog of the sum (\ref{terms-n>1}) with
\begin{equation}
\label{f-nonrel}
F(n) = 2 [n_F(M_n) + n_F(-M_n)]
\end{equation}
is diverging. To make the series convergent, one can introduce the regularization factor, $\exp(- \delta M_n)$ with $\delta \to +0$.
Then the use of the Euler-Maclaurin formula (\ref{E-M}) is justified. Taking into account the identity
$n_F(\epsilon) = 1/2[1 + \tanh (\mu-\epsilon)/2T]$, one can see that the field dependent densities coincide,
viz. $\rho_{\mathrm{I}}  = \rho^{\mathrm{rel}}_{\mathrm{I}}$ and $\rho_{\mathrm{II}} = \rho^{\mathrm{rel}}_{\mathrm{II}}$.

One can also come to the same conclusion that the magnetic field dependent parts of the carrier densities $\rho$
and $\rho^{\mathrm{rel}}$ coincide by using Eq.~(\ref{Omega-linked}). Indeed, taking the derivative over $\mu$,
one can see that
\begin{equation}
\label{density-link}
\rho^{\mathrm{rel}}(T,\mu,B) = \rho (T,\mu,B) - \frac{1}{2} \int_{-\infty}^{\infty} d\epsilon D(\epsilon).
\end{equation}
Here, in contrast to  Eq.~(\ref{Omega-linked-symmetric}) we do not assume the evenness of
the DOS $D(\epsilon)$. The integral in the RHS of Eq.~(\ref{density-link}) corresponds to
the total number of states in our bands which should be independent of the magnetic field
\cite{Tabert2015JPCM}.

\section{Magnetic field dependent part of the grand thermodynamic potential and magnetization}
\label{sec:magnetization}

Using Eq.~(\ref{Omega-integrate}) we restore  the thermodynamic potentials, viz.
\begin{equation}
\label{Omega-rel-restored}
\fl
\Omega^{\mathrm{rel}} (T,\mu,B) = \Omega^{\mathrm{rel}} (T,\mu,B=0)
 +  \Omega^{\mathrm{rel}}_{\mathrm{I}} (T,\mu,B) + \Omega^{\mathrm{rel}}_{\mathrm{II}} (T,\mu,B)
+ \Omega_c(T,B),
\end{equation}
and
\begin{equation}
\label{Omega-restored}
\Omega (T,\mu,B) = \Omega^{\mathrm{rel}} (T,\mu,B) +   \tilde \Omega_{c} (T,B).
\end{equation}
Here $\Omega^{\mathrm{rel}}_{\mathrm{I}}$ and  $\Omega^{\mathrm{rel}}_{\mathrm{II}}$
are, respectively, the linear and quadratic  in $B$ parts of the thermodynamic potential
obtained by integration of Eq.~(\ref{number-linear-B-fin}) for $\rho^{\mathrm{rel}}_{\mathrm{I}}$ and
Eq.~(\ref{number-quadratic-B-fin}) for $\rho^{\mathrm{rel}}_{\mathrm{II}}$ from $-\infty$ to $\mu$:
\begin{equation}
\label{Omega-1}
\Omega^{\mathrm{rel}}_{\mathrm{I}} (T,\mu,B)  =
- \frac{eB \mbox{sgn} (\eta \Delta) T }{4 \pi \hbar c}
\left[
\ln\cosh \frac{\mu + |\Delta|}{2T} - \ln\cosh \frac{\mu - |\Delta|}{2T}
\right]
\end{equation}
and
\begin{equation}
\label{Omega-2}
\Omega^{\mathrm{rel}}_{\mathrm{II}}  (T,\mu,B) = \frac{(eB)^2  v_F^2}{48 \pi c^2 |\Delta| }
\left[
\tanh \frac{\mu + |\Delta|}{2T} - \tanh \frac{\mu - |\Delta|}{2T}  \right] .
\end{equation}
In Eqs.~(\ref{Omega-rel-restored}) and (\ref{Omega-restored})
the arbitrary functions $\Omega_{c} (T,B)$ and $\tilde \Omega_{c} (T,B)$
may depend on $T$ and/or $B$, but independent of $\mu$.

Then the magnetization $M$ defined by Eq.~(\ref{magnetization-def}) in the low-field limit
reads
\begin{equation}
\label{magnetization-final}
M(T,\mu,B) = M_{\mathrm{I}} (T,\mu) +  M_{\mathrm{II}} (T,\mu,B) + M_c(T,B),
\end{equation}
where
\begin{equation}
\label{magnetization-B=0}
M_{\mathrm{I}} = - \frac{\partial  \Omega^{\mathrm{rel}}_{\mathrm{I}}}{\partial B}
= \frac{e \, \mbox{sgn} (\eta \Delta) T }{4 \pi \hbar c}
\left[
\ln\cosh \frac{\mu + |\Delta|}{2T} - \ln\cosh \frac{\mu - |\Delta|}{2T}
\right],
\end{equation}
is the anomalous magnetization,
\begin{equation}
\label{magnetization-B=1}
M_{\mathrm{II}}  = - \frac{\partial  \Omega^{\mathrm{rel}}_{\mathrm{II}}}{\partial B} =
- \frac{e}{24 \pi \hbar c} \frac{eB \hbar v_F^2}{c |\Delta| }
\left[
\tanh \frac{\mu + |\Delta|}{2T} - \tanh \frac{\mu - |\Delta|}{2T}  \right],
\end{equation}
is the linear in field part,
and the function $M_c(T,B)$ is independent of $\mu$. Our final result (\ref{magnetization-final})
with $M_{\mathrm{I}}$ and $M_{\mathrm{II}}$ given by Eqs.~(\ref{magnetization-B=0}) and (\ref{magnetization-B=1})
does not depend on a choice of the starting thermodynamic potential that may be taken either
(\ref{Omega.rel-def}) or (\ref{Omega.nonrel-def}).
The condition that at half-filling the anomalous magnetization is absent, viz. $M(T,\mu=0,B=0) = 0$ allows
to fix the function $M_c(T,B) =0$.

Persisting in the $B\to 0$ limit anomalous magnetization $M_{\mathrm{I}}$ was derived in
\cite{Gusynin2014PRB} using the low-field expansion for the Green's function in an external magnetic field.
This powerful and more complicated method allows one to consider interacting systems, while
the presented here method  uses the explicit form of the noninteracting spectrum (\ref{spectrum}).
In the $T=0$ limit Eq.~(\ref{magnetization-B=0}) acquires the form
\begin{equation}
M_{\mathrm{I}}=\frac{e}{4\pi\hbar c}
 \left[\eta \Delta \, {\rm sgn} \, (\mu ) \theta(|\mu|-|\Delta|) \right.
\left.
+\mu \, {\rm sgn} \, (\eta \Delta) \theta(|\Delta|-|\mu|)\right].
\end{equation}

For $\Delta_T =0$ and $\Delta_\eta = \Delta$, the anomalous $M_{\mathrm{I}}$ contribution from
$\mathbf{K}_{\pm}$ points has the opposite sign. Accordingly, when the integral contribution of
both valleys is considered,  the leading contribution to the magnetization is
linear in $B$ term given by Eq.~(\ref{magnetization-B=1}).
In the $T =0$ limit it reduces to
\begin{equation}
\label{magnetization-B-1-T=0}
M_{\mathrm{II}} = - \frac{e}{12 \pi \hbar c} \frac{eB \hbar v_F^2}{c |\Delta| } \theta(\Delta^2 - \mu^2), \quad L(B) \ll |\Delta|.
\end{equation}
This result is in agreement with the papers \cite{Koshino2010PRB,Sharapov2004PRB} where the gapped graphene was considered,
if one takes into account the factor of 4 from the spin-valley degeneracy.
In the limit $\Delta \to 0$ the corresponding magnetic susceptibility
was firstly considered by McClure \cite{McClure1956PRev} in the framework of the studies of
diamagnetism of graphite. In the opposite limit, $L(B) \gg |\Delta|$,
the vacuum term results in the magnetization $M(\mu=T=\Delta=0) \sim - \sqrt{B}$ (see \cite{Nersesyan1989JLTP,Schakel1991PRL,Raoux2014PRL,Sharapov2004PRB}
and references therein). Finally we note that
the impact of disorder on the magnetization and susceptibility of graphene
with $\Delta =0$ was studied in  \cite{Koshino2007PRB}.

\section{Hall conductivity}
\label{sec:Hall}

It is shown in \cite{Gusynin2014PRB} that the anomalous magnetization (\ref{magnetization-B=0}) is crucial
in obtaining the off-diagonal thermal transport coefficient. Below we illustrate how this magnetization can be
used to obtain the anomalous Hall conductivity and show how to obtain the Hall conductivity quantization in graphene
from Eq.~(\ref{number-rel1}).

\subsection{Anomalous quantum Hall effect}

When the chemical potential $\mu$ falls in a gap of the energy spectrum, the Hall conductivity can be found
from the St\v{r}eda formula \cite{Streda1982JPC}:
\begin{equation}
\label{Streda-formula}
\sigma_{xy} = -e c  \left( \frac{\partial \rho}{\partial B} \right)_\mu = -e c  \left( \frac{\partial M}{\partial \mu} \right)_B,
\end{equation}
where the second equality follows from the Maxwell relation.
Then either substituting the carrier density  (\ref{number-linear-B-fin}) in the first equality
of Eq.~(\ref{Streda-formula}) or substituting the  anomalous magnetization (\ref{magnetization-B=0}) in
the second equality of Eq.~(\ref{Streda-formula}), one obtains
\begin{equation}
\label{Hall-anomalous}
\sigma_{xy}(B=0)  = - \frac{e^2 \mathrm{sign} \, (\eta \Delta)}{8 \pi \hbar}
\left[\tanh \frac{\mu + |\Delta|}{2T} - \tanh \frac{\mu - |\Delta|}{2T} \right].
\end{equation}
We stress that this value of the anomalous Hall conductivity does not depend on a choice of the starting thermodynamic potential.
Furthermore, because Eq.~(\ref{Hall-anomalous}) corresponds to the second derivative of the
thermodynamic potential with respect to both $B$ and $\mu$, the presence of an arbitrary function of $T$ and $B$
in $\Omega$  and in the magnetization  (\ref{magnetization-final}) does not affect the final result.

In the limit $T=0$ the last equation reduces to
\begin{equation}
\label{Hall-anomalous-T=0}
\sigma_{xy}= - \frac{e^2}{4 \pi \hbar} \mathrm{sign} \, (\eta \Delta)  \theta (|\Delta| - |\mu| ).
\end{equation}
It is interesting to note that this result corresponds to the long-wavelength limit of the static Hall conductivity,
viz. $\lim_{q\to 0} \lim_{\omega \to 0} \sigma_{xy} (\omega,q)$ (see \cite{Ludwig1994PRB} and the discussion in
\cite{Koshino2011PRB}). The usual Hall conductivity, relevant for the transport, is given by an opposite limit,
$\sigma _{xy}^{\mathrm{tr}}= \lim_{\omega \to 0} \lim_{q\to 0} \sigma_{xy} (\omega,q)$.
This anomalous Hall conductivity was studied in \cite{Sinitsyn2006PRL,Sinitsyn2007PRB}
and in the clean limit for $\eta = +$ reads
\begin{equation}
\label{Hall-clean-final}
\sigma_{xy}^{\mathrm{tr}} =
-\frac{e^{2} \mathrm{sgn}\, (\Delta) }{4\pi \hbar }
\cases{
1, &  $|\mu | \leq |\Delta | ,$ \\
|\Delta| /|\mu | , & $|\mu | > |\Delta |.$ \\}
\end{equation}
We observe that Eq.~(\ref{Hall-clean-final}) agrees with Eq.~(\ref{Hall-anomalous-T=0}) when the chemical potential
is inside the gap, $|\mu| < |\Delta|$ in accord with the conditions of the St\v{r}eda formula \cite{Streda1982JPC} validity.

Finally we note that the presence of  disorder might regularize \cite{Ludwig1994PRB} the uncertainty of the order of limits mentioned below Eq.~(\ref{Hall-anomalous-T=0}). For  $|\mu | \leq |\Delta |$ the quantized value of the Hall conductivity (\ref{Hall-clean-final})
is robust with respect to disorder, while for $|\mu | > |\Delta |$ the presence of disorder modifies the result for the clean case
(see Refs.~\cite{Sinitsyn2006PRL,Sinitsyn2007PRB} and \cite{Ado2015} for a recent discussion).

\subsection{Quantum Hall effect in graphene}

So far we have considered the anomalous Hall effect by taking the limit $B=0$. To complete our consideration
is instructive to consider the Hall conductivity at finite field. Since the magnetization (\ref{magnetization-final})
is valid in the low-field regime, we come back to Eq.~(\ref{number-rel1}) for the carrier density, because it is valid for an arbitrary
strength of the magnetic field.
The same expression can be directly obtained from the Green's function
by doing the summation over Matsubara frequencies \cite{Niemi1985NPB,Lykken1990PRD,Schakel1991PRD}.
It is important that, in contrast to the the nonrelativistic case, the Matsubara summation in the relativistic case
is done \cite{Schakel1991PRD} without an additional convergence factor. When this factor is present, the Matsubara sum
produces the Fermi function $n_F$ as we had in Eq.~(\ref{number-nonrel-1}),
rather than $\tanh$ that is present in Eq.~(\ref{number-rel1}). Note also that to compare the corresponding to
Eq.~(\ref{number-rel1}) expression  from  Ref.~\cite{Schakel1991PRD}, one should rewrite the lowest Landau level
contribution $\rho^{\mathrm{rel}}_0$ as follows
\begin{equation}
\label{number-0}
\rho^{\mathrm{rel}}_0  =  \frac{eB}{4 \pi \hbar c}
\left[ \theta (eB)\tanh \frac{\mu + \eta \Delta }{2T} - \theta (-eB)\tanh \frac{\mu - \eta \Delta }{2T} \right],
\end{equation}
where we used that  $|eB| = eB \, \mbox{sgn} (eB) = eB [\theta(eB) - \theta(-eB)] $.
Using that
in the $T \to 0$ limit  $\tanh (\epsilon - \mu)/T  \to 1 - 2 \theta(- \epsilon+\mu)$, one can
show that Eq.~(\ref{number-0}) at $T=0$ acquires the form
\begin{equation}
\label{number-0-T=0-final}
\rho_0^{\mathrm{rel}} =  \frac{eB}{4 \pi \hbar c} \mbox{sgn} \,(\eta  \Delta) \theta (|\Delta| - |\mu|)
 + \frac{|e B|}{4 \pi \hbar c} \mbox{sgn} \, \mu \theta (|\mu| - |\Delta|).
\end{equation}
Similarly one can consider the $T=0$ limit for $\rho^{\mathrm{rel}}_{n \geq 1}$ part given
by Eq.~(\ref{terms-n>1}) and obtain
\begin{equation}
\label{number-n-T=0-final}
\rho_{n\geq 1} = \frac{|eB| \mbox{sgn} \, (\mu )}{ 2 \pi \hbar c} \sum_{n=1}^{\infty} \theta (|\mu| - M_n) .
\end{equation}
One can see that the first term of Eq.~(\ref{number-0-T=0-final})
corresponding to the anomalous Hall conductivity (\ref{Hall-anomalous})
is time-reversal symmetry breaking, since it is odd under $B \to - B$.
On the other hand, the second term of Eq.~(\ref{number-0-T=0-final}) and the whole Eq.~(\ref{number-n-T=0-final})
are even under $B \to - B$. For $\mu =0$ in the limit $\Delta \to 0$ the first term of Eq.~(\ref{number-0-T=0-final})
leaves behind the $\mbox{sgn} \,(\eta  \Delta)$. This property is called a sign anomaly. As noted in \cite{Niemi1985NPB},
this anomaly is removed by a finite density effects. Indeed, for $\mu \neq 0$ only the second term
of Eq.~(\ref{number-0-T=0-final}) survives in the limit $\Delta \to 0$.

Adding together $\rho_0^{\mathrm{rel}}$ and $\rho_{n\geq 1}^{\mathrm{rel}}$ we recover the results of Lykken {\em et al.} \cite{Lykken1990PRD}
and Schakel \cite{Schakel1991PRD}
\begin{equation}
\label{Schakel-imbalance}
\rho^{\mathrm{rel}} =  \frac{eB}{4 \pi \hbar c} \mbox{sgn} \,(\eta  \Delta) \theta (|\Delta| - |\mu|)
 + \frac{|eB| \mbox{sgn} \, (\mu)}{ 2 \pi \hbar c} \left( \frac{1}{2}+
\left[ \frac{(\mu^2 - \Delta^2) c}{2 \hbar v_F^2 |e B| }\right] \right),
\end{equation}
where $[x]$ denotes the integer part of $x$. As discussed above, graphene corresponds to the case with
$\Delta_T =0$, $\Delta_\eta = \Delta \to 0$. Taking into account the spin-valley degeneracy and
using the St\v{r}eda formula (\ref{Streda-formula}) we reproduce
the half-integer quantum Hall effect \cite{Ando2002,Gusynin2005PRL,Peres2006PRB}
\begin{equation}
\label{QHE-graphene}
\sigma_{xy} =- \frac{2 e^2 \mbox{sgn} \, (eB) \mbox{sgn} \, (\mu)}{\pi \hbar}
\left( \frac{1}{2} +
\left[ \frac{\mu^2 c}{2 \hbar v_F^2 |e B| }\right]  \right).
\end{equation}
The argument given below Eq.~(\ref{density-link}) allows to conclude that
the same result also follows from the nonrelativistic carrier density.
The Hall conductivity (\ref{QHE-graphene}) can also be written as
a function of the free carrier imbalance, $\rho_{\mathrm{free}}^{\mathrm{rel}}$,
\begin{equation}
\sigma_{xy} =- \frac{2 e^2 \mbox{sgn} \, (eB) \mbox{sgn} \, (\mu)}{\pi \hbar}
\left( \frac{1}{2} +
\left[ \frac{\pi \hbar c \rho_{\mathrm{free}}^{\mathrm{rel}} }{2 |e B| }\right]  \right),
\end{equation}
where we used that in the absence of magnetic field
$\rho_{\mathrm{free}}^{\mathrm{rel}} = \mu^2 \mbox{sgn} \, (\mu)/(\pi \hbar^2 v_F^2)$.
Concluding it is worth to stress that the quantum Hall effect is a disorder-induced phenomenon. Accordingly
the presented consideration only illustrates the expected quantization of the Hall conductivity, rather than explains it.

\section{Conclusion}
\label{sec:concl}

In the present work we demonstrated how the Landau approach used to describe the diamagnetism of electron gas
can be applied for the $2+1$-dimensional massive Dirac fermions. We derived an explicit expression
(\ref{magnetization-final}) for magnetization in the low-field regime. It contains
field independent (anomalous) term (\ref{magnetization-B=0}) and the linear in field part
(\ref{magnetization-B=1}). While there exist many approaches that allow one to obtain these terms,
the use of the Euler-Maclaurin formula as in the original work Landau \cite{Landau1930ZPhys} is the simplest option.
Using the St\v{r}eda formula it is illustrated how the obtained expressions for the carrier imbalance (\ref{number-linear-B-fin}) (or Eq.~(\ref{magnetization-B=0}) for the magnetization) and (\ref{Schakel-imbalance}) are related to
the anomalous quantum Hall effect and quantum Hall effect in graphene.

\ack
S.G.Sh gratefully acknowledges E.V.~Gorbar,
V.P.~Gusynin, V.Yu. Tsaran, and A.A.~Varlamov for helpful discussions.


\section*{References}

\begin{thebibliography}{99}

\bibitem{Semenoff1984PRL} Semenoff G W 1984 {\it Phys. Rev. Lett.} \textbf{53} 2449

\bibitem{Haldane1988PRL} Haldane F D M 1988 {\it Phys. Rev. Lett.} \textbf{61} 2015

\bibitem{Geim2007NatMat} Geim A K and Novoselov K S 2007 {\it Nature Materials} \textbf{6} 183

\bibitem{Davila2014NJP} D\'{a}vila M E, Xian L, Cahangirov S, Rubio A and Le~Lay G 2014 {\it New J. Phys.} {\bf 16} 095002

\bibitem{Gomes2012Nature} Gomes K K, Mar W, Ko W, Guinea F and Manoharan H C 2012, {\it Nature} \textbf{483} 306

\bibitem{Tarruell2012Nature} Tarruell L, Greif D, Uehlinger T, Jotzu G and Esslinger T 2012
{\it Nature} \textbf{483} 302

\bibitem{Jotzu2014Nature} Jotzu  G, Messer M, Desbuquois R, Lebrat M, Uehlinger T, Greif D and Esslinger T 2014
{\it Nature} \textbf{515} 237

\bibitem{Nielsen1981NPB} Nielsen H B and Ninomiya M 1981 {\it Nucl. Phys.}  \textbf{B185} 20;
Nielsen H B and Ninomiya M  1981 {\it Nucl. Phys.}  \textbf{B193} 173

% our review
\bibitem{Gusynin2007IJMPB} Gusynin V P, Sharapov S G and Carbotte J P 2007
{\it Int. J. Mod. Phys. B} \textbf{21} 4611

\bibitem{Tchernyshyov2000PRB} Tchernyshyov O 2000 {\it Phys. Rev. B} {\bf 62} 16751

\bibitem{Hosotani1993PLB} Hosotani Y 1993 {\it Phys. Lett. B} \textbf{319} 332;  Hosotani Y
1995 {\it Phys. Rev. D} \textbf{51} 2022

\bibitem{Andresen1995PRD} Andersen J O and Haugset T
1995 {\it Phys. Rev. D} \textbf{51} 3073

\bibitem{Cangemi1996AP} Cangemi D and Dunne G 1996 {\it Ann. Phys.} \textbf{249} 582

\bibitem{Sharapov2004PRB} Sharapov S G, Gusynin V P and Beck H 2004
{\it Phys. Rev. B} {\bf 69} 075104

\bibitem{Tabert2015PRB} Tabert C J, Carbotte J P and Nicol E J 2015  {\it Phys. Rev. B}  \textbf{91} 035423

\bibitem{Landau5:book}  Landau L D and Lifshitz E M 1980 {\it Statistical Physics. Vol. 5 } (3rd ed.) (Oxford: Pergamon Press)

\bibitem{Niemi1985NPB} Niemi A J 1985 {\it Nucl. Phys. B} \textbf{251} [FS13] 155

\bibitem{Gusynin1995PRD} Gusynin  VP, Miransky V A and  Shovkovy I A 1994 {\it Phys.~Rev.~Lett.} {\bf 73} 3499;
Gusynin  V P, Miransky V A and  Shovkovy I A 1995 {\it Phys.~Rev.~D} {\bf 52} 4718

\bibitem{Niemi1986PR} Niemi A J and Semenoff G W 1986 {\it Phys. Rep.} \textbf{135} 99

\bibitem{Landau1930ZPhys} Landau L 1930 {\it Z. Phys.} \textbf{64} 629

\bibitem{Tabert2015JPCM} Tabert C J and Carbotte J P 2015
{\it J. Phys.: Condens. Matter} \textbf{27} 015008

\bibitem{Gusynin2014PRB} Gusynin V P,  Sharapov S G and Varlamov A A 2014 {\it Phys. Rev.  B} {\bf 90} 155107

\bibitem{Koshino2010PRB} Koshino M and Ando T 2010 {\it Phys. Rev.  B} {\bf 81} 195431;
Koshino M and Ando T 2011 {\it Sol. St. Commun.} \textbf{151} 1054

\bibitem{McClure1956PRev} McClure J W 1956   {\it Phys. Rev.} \textbf{104} 666;
McClure J W 1960 {\it Phys. Rev.} \textbf{119} 606

\bibitem{Nersesyan1989JLTP}  Nersesyan A A and Vachanadze G E 1989 {\it J. Low Temp. Phys.} \textbf{77} 293

\bibitem{Schakel1991PRL} Schakel A M J and Semenoff G W 1991 {\it Phys. Rev. Lett.} \textbf{66} 2653

\bibitem{Raoux2014PRL} Raoux A, Morigi M, Fuchs J-N, Pi\'{e}chon F and Montambaux G 2014
{\it Phys. Rev. Lett.} \textbf{112}  026402

\bibitem{Koshino2007PRB} Koshino M and Ando T 2007 {\it Phys. Rev.  B} {\bf 75} 235333


% Streda
\bibitem{Streda1982JPC} St\v{r}eda P 1982 {\it J. Phys. C} {\bf 15} L717

\bibitem{Ludwig1994PRB} Ludwig A W W, Fisher M P A, Shankar R and Grinstein G
1994 {\it Phys. Rev. B} \textbf{50} 7526

\bibitem{Koshino2011PRB} Koshino M 2011 {\it Phys. Rev.  B} {\bf 84} 1125427

\bibitem{Sinitsyn2006PRL} Sinitsyn N A, Hill J E, Min H, Sinova J and
MacDonald A H  2006 {\it Phys. Rev. Lett.} \textbf{97} 106804

\bibitem{Sinitsyn2007PRB} Sinitsyn N A, MacDonald A H, Jungwirth T, Dugaev V K and Sinova J
2007 {\it Phys. Rev. B} \textbf{75} 045315

\bibitem{Ado2015} Ado A, Dmitriev I A, Ostrovsky P M and Titov M 2015 Preprint arXiv:1504.03658

% relativisitic
\bibitem{Lykken1990PRD} Lykken J D, Sonnenschein J and Weiss N
1990 {\it Phys.~Rev.~D} {\bf 42} 2161

\bibitem{Schakel1991PRD} Schakel A M J 1991 {\it Phys.~Rev.~D} {\bf 43} 1428

\bibitem{Ando2002} Zheng Y and Ando T 2002 {\it Phys. Rev. B} {\bf 65} 245420

\bibitem{Gusynin2005PRL} Gusynin  V P and Sharapov S G 2005
{ \it Phys. Rev. Lett.} {\bf 95} 146801;
Gusynin  V P and Sharapov S G 2006 {\it Phys. Rev. B} {\bf 73} 245411

\bibitem{Peres2006PRB} Peres N M R, Guinea F and Castro Neto A H  2006 {\it Phys. Rev. B} {\bf 73} {125411}




\end{thebibliography}
\end{document}